\documentclass[journal=jacsat,manuscript=article,layout=twocolumn,email=true]{achemso}

\usepackage[version=3]{mhchem} % Formula subscripts using \ce{}
\usepackage{epstopdf}  % Nice! this epstopdf package
\usepackage{mciteplus}
\usepackage{fixltx2e}
\author{Guohui Hu}
\affiliation[Shanghai University]{Shanghai Institute of Applied Mathematics and Mechanics, \\ Shanghai Key Laboratory of Mechanics in Energy Engineering, Shanghai University, \\ 149 Yanchang Road, Shanghai 200072, P. R. China}
\alsoaffiliation[Northwestern University]{ \\ Department of Mechanical Engineering, Northwestern University, \\ 2145 Sheridan Rd, Evanston, IL, 60208, USA}
\author{Mao Mao}
\affiliation[Northwestern University]{ \\ Department of Mechanical Engineering, Northwestern University, \\ 2145 Sheridan Rd, Evanston, IL, 60208, USA}
\author{Sandip Ghosal}
\affiliation[Northwestern University]{ \\ Department of Mechanical Engineering, Northwestern University, \\ 2145 Sheridan Rd, Evanston, IL, 60208, USA}
\email{s-ghosal@northwestern.edu}

\title{Ion transport through a graphene nanopore}
\begin{document}

%\author{Guo-Hui Hu\textsuperscript{1,2}, Mao Mao\textsuperscript{2}, Sandip
%Ghosal\textsuperscript{2}%
%\thanks{Corresponding author. Email: s-ghosal@northwestern.edu.%
%}}
%
%\maketitle
%
%\lyxaddress{\textsuperscript{1}Shanghai Institute of Applied Mathematics and
%Mechanics, Shanghai Key Laboratory of Mechanics in Energy Engineering,
%Shanghai University, 149 Yanchang Road, Shanghai 200072, P. R. China }
%
%
%\lyxaddress{\textsuperscript{2}Department of Mechanical Engineering, Northwestern
%University, 2145 Sheridan Road, Evanston, IL 60208, USA }

\begin{abstract}
Molecular dynamics simulation is utilized to investigate the ionic
transport of NaCl in solution through a graphene nanopore under an
applied electric field. Results show the formation of concentration
polarization layers in the vicinity of the graphene sheet. The non-uniformity
of the ion distribution gives rise to an electric pressure which drives
vortical motions in the fluid if the electric field is sufficiently
strong to overcome the influence of viscosity and thermal fluctuations.
The relative importance of hydrodynamic transport and thermal fluctuations
in determining the pore conductivity is investigated. A second important
effect that is observed is the mass transport of water through the
nanopore, with an average velocity proportional to the applied voltage
and independent of the pore diameter. The flux arises as a consequence
of the asymmetry in the ion distribution with respect to reflection about the plane of the graphene sheet. The accumulation of liquid molecules
in the vicinity of the nanopore due to reorientation of the water
dipoles by the local electric field is seen to result in a local increase
in the liquid density. Results confirm that the electric conductance
is proportional to the nanopore diameter for the parameter regimes
that we simulated. The occurrence of fluid vortices is found to result
in an increase in the effective electrical conductance. 
\end{abstract}

\section{Introduction}

Ionic conduction through nanometer sized channels or pores is a common
theme in biological systems as well as in various manufactured materials
such as membranes and synthetic nanopores \cite{doyle1998thestructure,rhee2006nanopore,venkatesan2011nanopore,branton2008thepotential,roux2004theoretical}.
Theoretical as well as experimental aspects of the problem have attracted
increasing interest in the past decade. Molecular dynamics (MD) simulation
is an effective tool for exploring these nanoscale phenomena. It has
the advantage of being able to relate the observables directly to
the molecular properties of the solid and liquid, once an appropriate
intermolecular potential is given, without the need for too many simplifying
assumptions. Various aspects of the problem, such as ionic current
rectification, DNA translocation and water transport have already
been reported in the literature \cite{aksimentiev2009modeling,aksimentiev2005imaging,comer2009microscopic,cruz-chu2009ioniccurrent,sathe2011computational,suk2010watertransport}.

MD simulations in the context of translocation of single stranded
and double stranded DNA through biological $\alpha$-Hemolysin and
synthetic nanopores have been reported \cite{aksimentiev2005imaging,comer2009microscopic,aksimentiev2004microscopic}.
It was found that the open-pore current increases linearly with the
applied voltage, and, obeys Ohm's law for voltages that are no more
than of the order of a Volt. The distribution of the electric potential
around the pore, the translocation speed of DNA, the interactions
between the DNA molecule with the pore wall, as well as strategies
for controlling the transolcation speed have been considered~\cite{luan2010control,mirsaidov2010slowing}.
Recently, Sathe et al. \cite{sathe2011computational} investigated
the translocation of DNA through a graphene nanopore using MD simulations,
and suggested that nucleotide pairs can be discriminated using graphene
nanopores under suitable bias conditions.

There are, however, a number of open problems that have not been understood.
First, due to the geometry of the system, the ions may accumulate
and form a concentration polarization layer (CPL) in the vinicity
of the membrane. The influence of the nonuniformity of the ion concentration
in this charge separated Debye layer on the ionic current has not
been well understood. Secondly, open questions remain on the effect
of hydrodynamic flow on the ionic current as well as on DNA translocation
speeds. Ghosal~\cite{ghosal_PRE06,ghosal_PRL07} presented a simple
hydrodynamic calculation for the electrophoretic speed of the polymer,
modeled locally as a long cylindrical object centered on the axis
of the pore. The calculation yielded analytical results in close agreement
with experimental measurements~\cite{storm_physRevE05,dekker_nano_lett06}.
In a related problem where the electrophoretic force was measured
with the DNA immobilized in the pore, numerical~\cite{van_dorp_origin_2009}
as well as analytical~\cite{ghosal_PRE07} models based on the hypothesis
that the hydrodynamic drag was the primary resistive force yielded
results in close agreement to the experiments. These results point
to the possibility that hydrodynamics might play an important role
in determining DNA translocation speeds. Hydrodynamics also plays
an important role in a related problem; when a direct current is applied
across an ion-selective nanoporous membrane or through a nanochannel
with overlapping Debye layers, it is known, that, micro fluid vortices
may be observed due to hydrodynamic instability, and these vortices
are capable of enhancing the ionic current in the so called ``overlimiting
regime'' \cite{chang2012nanoscale}.

The purpose of the present study is to discuss the influence of the
CPL and fluid convection on ionic transport through nanochannels using
MD simulations. The rest of this paper is organized as follows: in
Sec. 2, we present the physical model and describe the numerical approach.
The simulation results are analysed in detail in Sec. 3. Finally,
some concluding remarks are made in Sec. 4.

\section{Method}

\subsection{System Setup}

The molecular simulations are conducted in a cubic box with dimensions
of $L_{x}=5.112$ nm, $L_{y}=5.184$ nm, $L_{z}=10$ nm. The origin
of the coordinates is set in the center of the box. The graphene sheet
with the size of $L_{x}\times L_{y}$, consists of an array of carbon
atoms with a planar hexagonal structure (neighboring atoms have a
bond lengths of $0.142$ nm), localized at the mid-plane, $z=0$.
A nanopore, with radius $a$, is constructed by removing the carbon
atoms in a central circular patch of the graphene sheet: $x^{2}+y^{2}<a^{2}$.
The remainder of the box is then filled with water molecules described
by the extended simple point charge (SPC/E) model \cite{berendsen1987themissing}.
Polarization of water molecules is neglected, prior investigations
have shown that this is reasonable as long as the field strength does
not exceed about 10 V/nm \cite{yang2002canonical}. Salt (NaCl) is
introduced at a given concentration of $1$M by replacing the required
number of water molecules with Na$^{+}$ and Cl$^{-}$ ions.

\subsection{Molecular Dynamics (MD) simulation}

The simulations are performed at a constant temperature 300K and pressure
of 1 bar with the large scale MD package GROMACS 4.5.4 \cite{hess2008gromacs}.
The van der Waals (vdW) interaction of the carbon atoms is modeled
as uncharged Lennard-Jones (LJ) particles. The graphene-water interaction
is considered by a carbon-oxygen LJ potential. This general set up
and parameter values have been employed in previous studies \cite{gong2007achargedriven,hummer2001waterconduction,xiao-yan2007thestructure}.

A uniform external electric field is applied in directions perpendicular
to the graphene sheet ($z$-direction). The contribution of the external
electric field is described as $U_{E}=-{\displaystyle \sum q_{i}\mathbf{r}_{i}\cdot\mathbf{E}}$,
where the $q_{i}$ and $\mathbf{r}_{i}$ denote the charge and location
of the charged atom $i$ respectively, and $\mathbf{E}$ is the strength
of the external electric field. The LJ interactions are truncated
at the cut-off distance $r_{0}=1.0$ nm and the particle mesh Ewald
(PME) method \cite{darden1993particle} with a real-space cut-off
of $1$ nm is utilized to treat the long-range electrostatic interactions.
Periodic boundary conditions are imposed in all directions. The time
step in all simulations is set to be $2$ fs.

For the sake of computational efficiency, all of the carbon atoms
are frozen during the simulations. Previous investigations have shown
that this only has a minor influence on the dynamics of the adjacent
water. Minimization of energy is performed with the steepest descent
method on the initial system. Then the system is evolved for $2$
ns to achieve a state of statistical equilibrium. In all cases, statistics
are collected during the last $8$ ns and samples are taken every
$0.2$ ps.

\subsection{Data analysis}

To analyse the statistics of the macroscopic physical variables, the
spatial location and velocity vectors of the particles are transformed
to cylindrical coordinates since averages of variables should be close
to axisymmetric. Data is presented in the $r-z$ plane which is partitioned
into gridded cells with interval range ($\triangle r$,$\triangle z$
).

\textsc{Ionic concentration.} The number density of Na and Cl ions in
the cells are calculated by $\rho_{i}=N_{c}/V_{c}$, where $N_{c}$
and $V_{c}$ are the number of ions and volume of the corresponding
cell respectively.

\emph{\textsc{Water flux}}\emph{.} The number of oxygen atoms $N_{w}$
through the nanopore in time interval $\triangle t$ is calculated
to obtain the water flux. The average water velocity $v_{p}$ in the
axial direction through the nanopore is given by $v_{p}=mN_{w}/\triangle t\rho_{0}A$,
in which $m$ is the mass of a water molecule, $\rho_{0}$ is water
density in the bulk and $A=\pi a^{2}$ is the area of the nanopore.

\begin{figure}[!ht]
\centering
\includegraphics[width=0.5\textwidth]{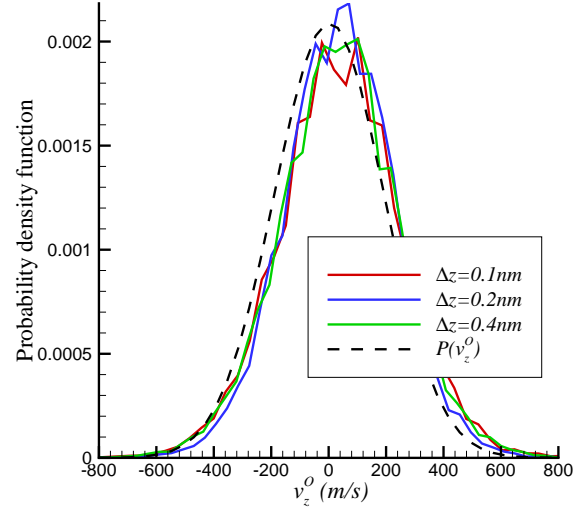}
\centering
\caption{Probability distribution function of $z$-component of flow velocity
at the nanopore for different sizes of grid. The velocity
distribution can be well represented by a Gaussian. }
\label{pdf}
\end{figure}

\textsc{Flow field.} For the cell at ($r$, $z$) containing $N_{c}$
water molecules, the transient local velocity vector can be obtained
by

\[
\mathbf{v}(r,z,t_{j})=\frac{1}{\triangle tN_{c}}\sum_{i=1}^{N_{c}}\left[\mathbf{r}_{Oi}(t_{j}+\triangle t)-\mathbf{r}_{Oi}(t_{j})\right]
\]
 where $\mathbf{r}_{Oi}(t_{j})$ is the location vector of oxygen
atom $i$ at time $t_{j}$. $\triangle t$ is the time interval between
two successive frames. We take $\triangle t=2$ ps in the present
study. The velocity field $\mathbf{\bar{v}}(r,z)$ of the equilibrium
state is then calculated by averaging the transient velocity over
time. To check the independence of physical quantities with respect
to grid size, the probability density function (PDF) of the axial
velocity $v_{z}^{O}$ of water at (0, 0) is plotted in Fig. \ref{pdf} for
different grid sizes. After determining the average velocity $\bar{v}_{z}^{O}$
and standard error $\sigma$, it is seen that the PDF is well represented
by a Gaussian:

\[
P(v_{z}^{O})=\frac{1}{\sigma\sqrt{2\pi}}\exp[-\frac{(v_{z}^{O}-\bar{v}_{z}^{O})^{2}}{2\sigma^{2}}]
\]
 Furthermore, the PDFs are nearly independent of the computational
grid size. In the present study, we set $\triangle z=0.2$ nm.

\textsc{Ionic current.} The ionic current is obtained by $I=eN_{i}/\Delta t$,
where $N_{i}$ stands for the number of Na or Cl ions across the graphene
nanopore in time interval $\Delta t$, $e$ is the proton charge.

\section{Results and Discussions}

\subsection{Concentration polarization}

\begin{figure*}[t]
\centering
\includegraphics[width=\textwidth]{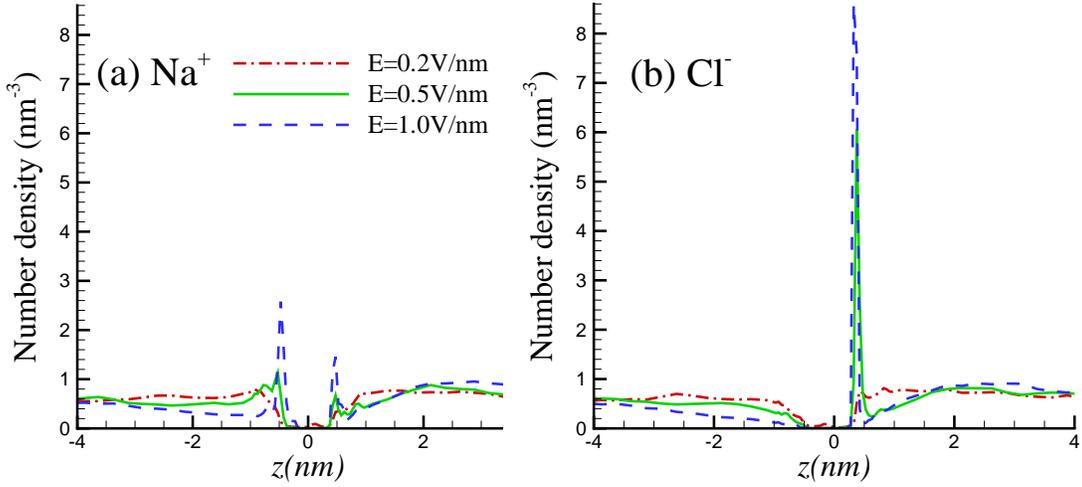}
\caption{Number density profile of Na and Cl ions in $z$-direction with $d$=1.5 nm
for different field strengths, showing the presence of concentration
polarization layer.}
\label{ion-z}
\end{figure*}

If the pore size approaches zero, clearly a polarized double layer
would form adjacent to the graphene sheet. For a finite, but nevertheless
sufficiently small pore size, a nonuniform ion distribution (concentration
polarization) is expected adjacent to the sheet. The profiles of ionic
number density along the $z$-axis is plotted in Fig. \ref{ion-z}, where $z=0$
corresponds to the graphene sheet. It is interesting that the sodium
and chlorine ions both form concentration polarization layers (CPL)
on either side of the graphene sheet. However, their distributions
show some asymmetry. Due to size exclusion, the ions cannot approach
the graphene sheet closer than about an ionic radius. The combination
of electrostatic and these steric forces result in the appearance
of a concentration peak. In the case of the chlorine ions, the concentration
peak is located a distance $D_{cl}=0.375$ nm from the sheet. This
distance is found to be independent of the applied field strength.
The CPL can also be observed for the sodium ions on the other side;
however, the amplitude of the peak is comparatively weak. For weak
applied fields (e.g. $E=0.2$ V/nm) the CPL is not as well defined.
A second smaller peak of sodium ions may be observed on the other
side of the membrane. The asymmetry between the two kinds of ions
may be partly due to the higher mobility of the chlorine ions, and,
partly due to differences in the van der Waals interactions between
the two kinds of ions with the carbon atoms in the graphene sheet.
The peak in the ionic concentration profile is found to increase monotonically
with increasing electric field.

%\afterpage{\clearpage}

\begin{figure*}
\centering
\includegraphics[scale=0.5]{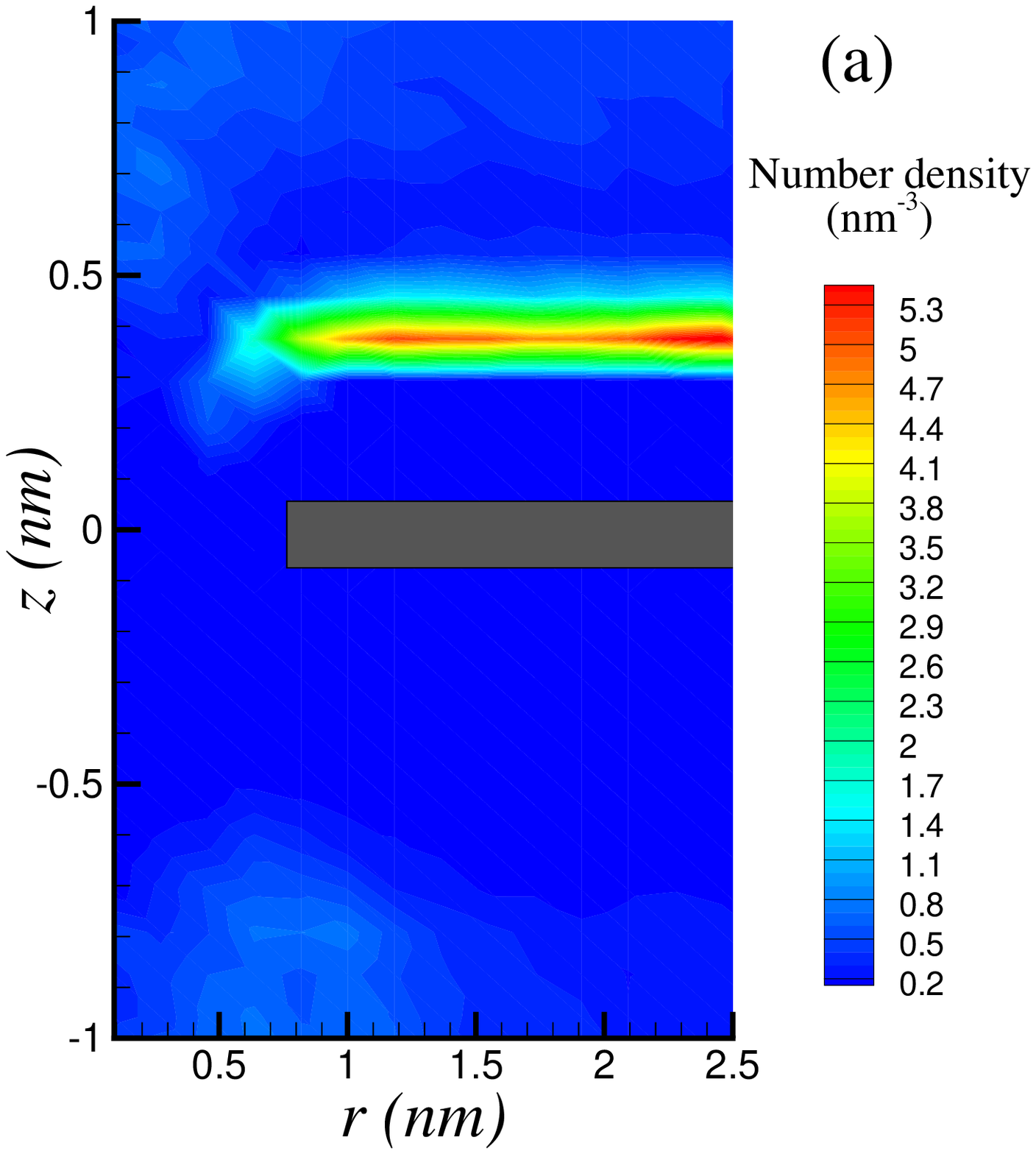}
\centering
\includegraphics[scale=0.5]{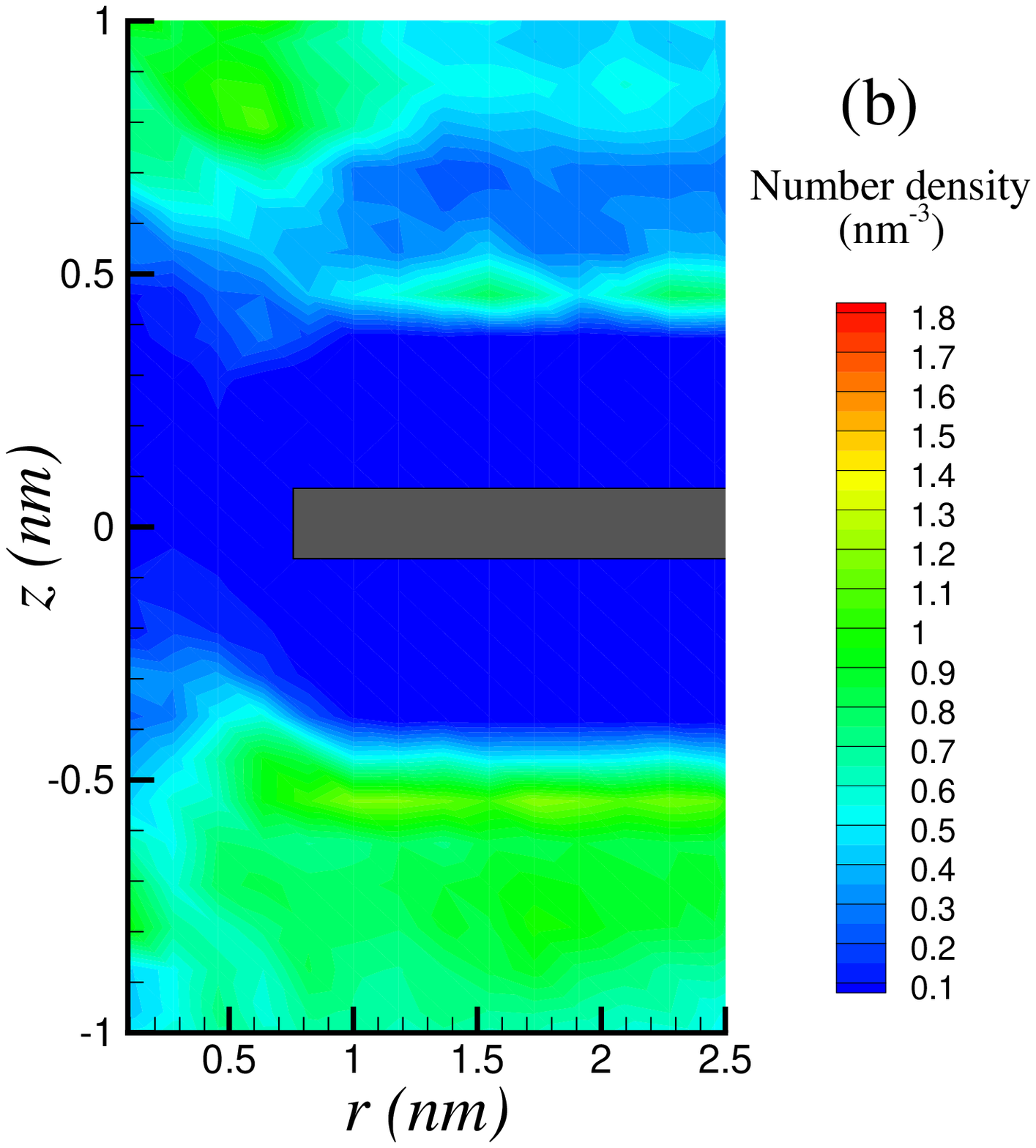}
\caption{Nonuniform distributions of number density of chlorine (a)
and sodium (b) ion with $d$=1.5 nm and $E$=0.5 V/nm. The
grey rectangle at $z$=0 marks the graphene sheet. Ionic gradients
lead to electric pressure, which drives nanoscale vortices near the
pore. }
\label{distribution}
\end{figure*}

In Fig. \ref{distribution}, the number density of Cl ions with $E=0.5$ V/nm and nanopore
diameter $d=1.5$ nm is observed to exhibit a well defined peak. The
nonuniformity of the ionic distribution will give rise to spatial
gradients of the electric field strength near the membrane, especially
in regions close to the nanopore. Consequently, variations of the
electric pressure is to be expected, which might drive a flow, as
long as these forces are strong enough to overcome viscous resistance
and are not completely masked by the fluctuating Brownian forces (thermal
fluctuations).

\subsection{Flow fields }

\begin{figure*}[!htp]
\centering
\includegraphics[scale=0.4]{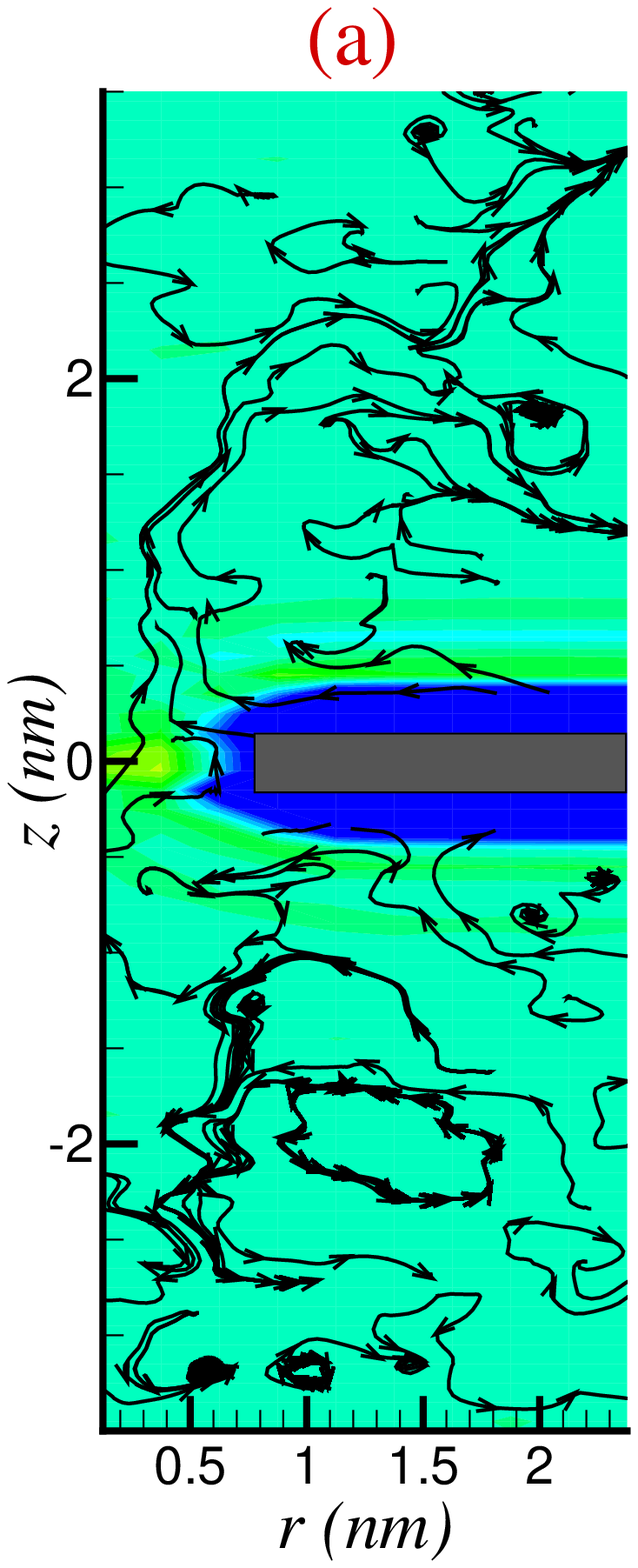}
\includegraphics[scale=0.4]{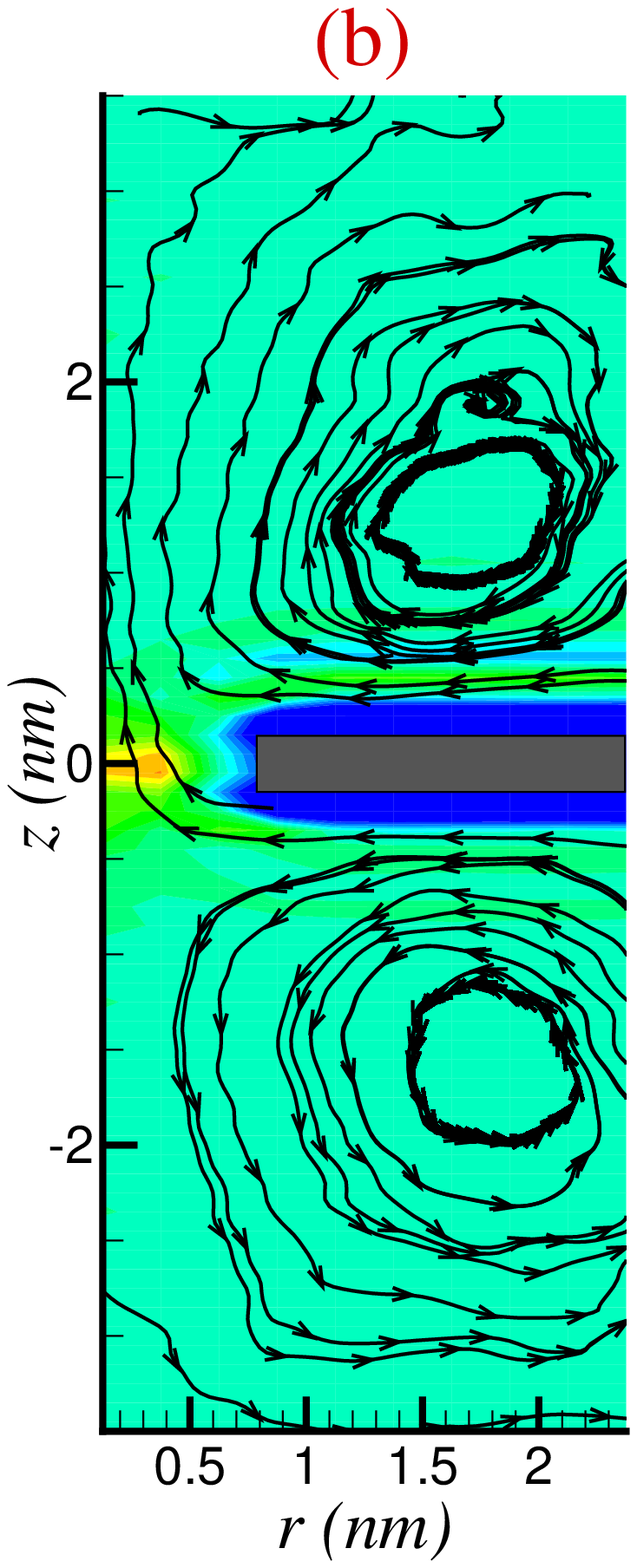}
\includegraphics[scale=0.4]{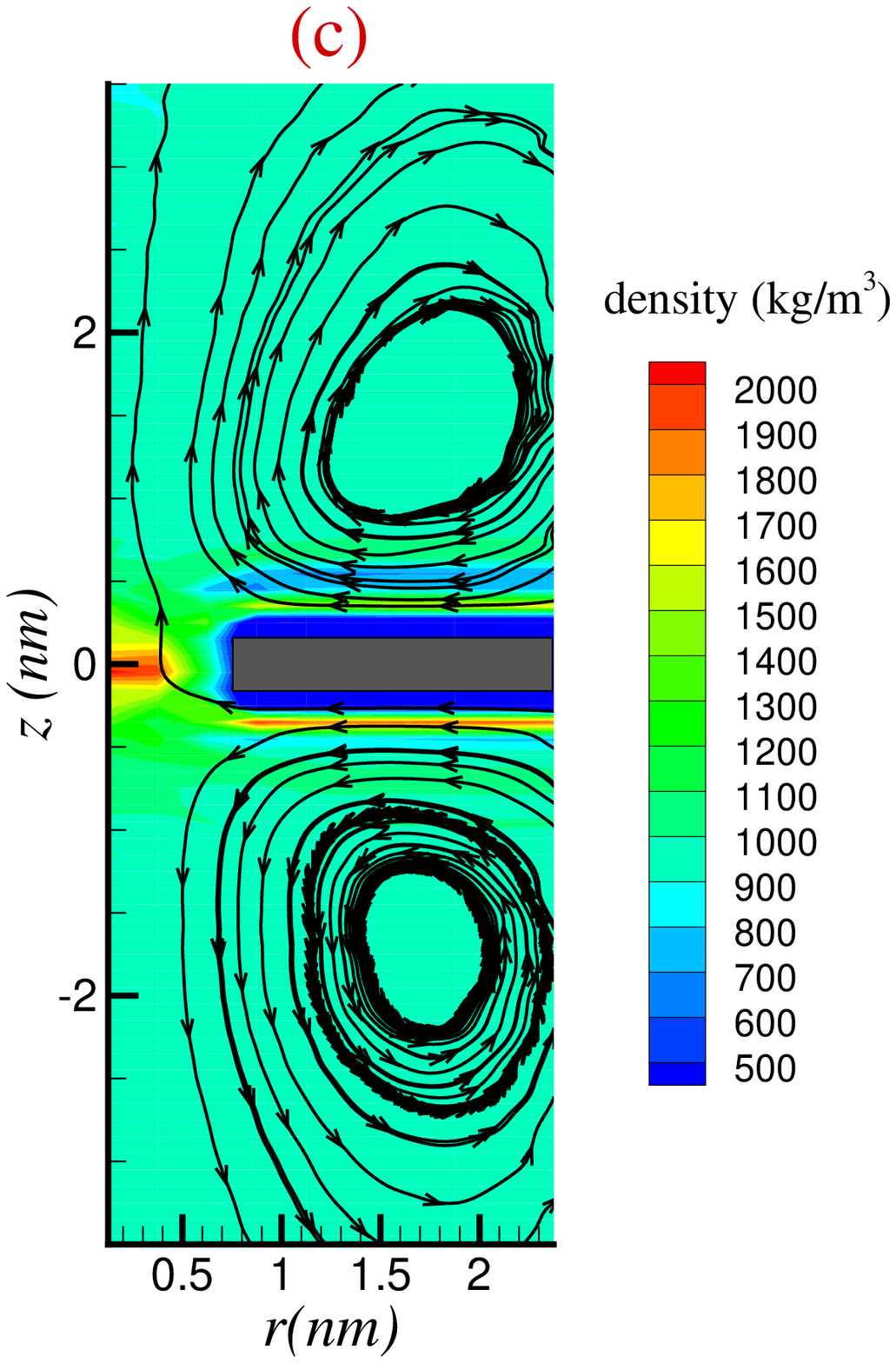}
\caption{Flow fields and water density (in units of kg/m$^{3}$) distribution
of salt solution for a nanopore of diameter $d$=1.5 nm with (a) $E$=0.2
V/nm; (b) $E$=0.5 V/nm; (c) $E$=1.0 V/nm. The rectangles (color
grey near $z$=0) marks the graphene sheet. For strong fields, as
in panel (c), a well defined vortical flow is seen but for weak electric
fields the bulk motion of the fluid is difficult to distinguish from
the background thermal fluctuations.}
\label{vel}
\end{figure*}

The formation of the CPL is expected to generate a fluid flow in the
vicinity of the interface. This is indeed seen in the simulations.
The flow streamlines for the nanopore of diameter $d=1.5$ nm, together
with the distribution of water density is shown in Fig. \ref{vel} with different
values of electric field strength. When the field is sufficiently
large, e.g., $E=0.5$ and $1.0$ V/nm, vortices of a spatial scale
on the order of nanometers are clearly observed. When $E$=0.2 V/nm,
the electric pressure appears too weak to generate sustained micro
vortices against the influence of thermal fluctuations.

\begin{figure}[!h]
\centering
\includegraphics[scale=0.4]{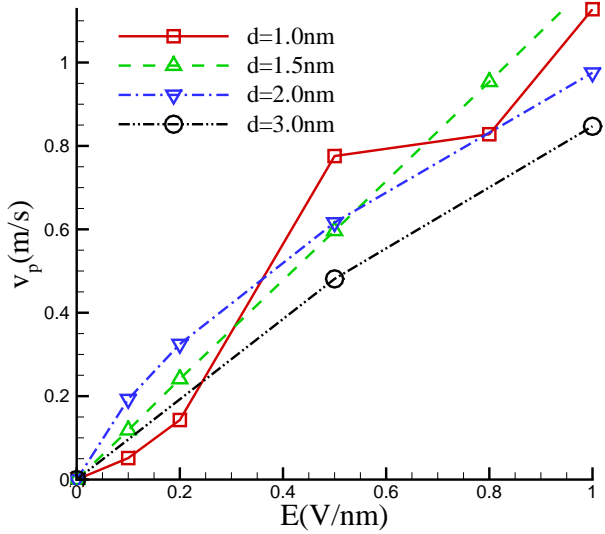}
\caption{Variation of the mean flow ($v_{p}$) at the nanopore with the applied
electric field for different pore diameters.}
\label{flux}
\end{figure}

An interesting physical insight that emerges from these simulations
is that there appears to be a marked asymmetry in the distribution
of the cations and anions and that of water density. This is most
likely due to a combination of three factors. First, as we have mentioned
above, there is an approximately $30$ percent difference in mobility
between sodium and chlorine ions. Second, the van der Waals interactions
between Na and C, Cl and C differ significantly. The third, which
has been reported by previous researchers \cite{bratko2007effectof,daub2006electrowetting,daub2012nanoscale,hu2008dewetting},
is that the direction of the electric field could have significant
effects on the interfacial water structures, as well as on the wetting
behavior of the solid membrane, due to the polarity of the water molecules.
A consequence of this asymmetry, is that there is a flux of water
through the nanopore as may be seen in the appearance of the streamlines
in Fig. \ref{vel}. The average flow velocity $v_{p}$, describing water crossing
the nanopore, is presented in Fig. \ref{flux} for different electric field
strengths and pore diameters. It shows that generally the average
velocity at the nanopore is proportional to the applied voltage, and
nearly independent of the diameter of the nanopore for the parameters
we simulated.

\begin{figure}[!h]
\centering
\includegraphics[scale=0.4]{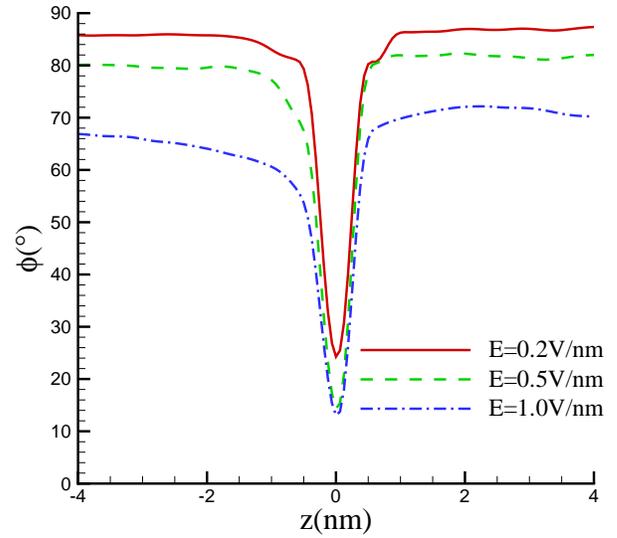}
\caption{Average angle between the dipole moment and the $z$-axis for a pore
diameter of $d=1.5$ nm. Reduction of the angle due to the re-orientation
of the dipole moment by the field is strongest in the pore region
where the dipoles favor to be parallel to the graphene sheet.}
\label{dipole}
\end{figure}

A region of high water density is seen at the center of the nanopore.
For $E=1.0$ V/nm, the local water density can be as high as $2043$
kg/m$^{3}$. The accumulation of water in the nanopore appears to
originate from re-orientation of the water dipole moments when the
external voltage is applied. To describe the orientation of water
molecules, we define an average over the $x$-$y$ plane of the angle
$\phi$ between the dipole moment vector of the water molecule and
the $z$-axis. In the absence of an electric field, the orientation
of the dipoles is completely random corresponding to $\phi=90^{\circ}$.
Fig. \ref{dipole} shows that with increasing electric field strengths, the dipoles
prefer an axial orientation. For a given $z$-location, this effect increases
with the field strength. The degree of ordering also increases as one approaches
the pore where the dipoles prefer to be parallel to the graphene surface. This is consistent with previous studies on SWCNT where
a similar ordering has been observed \cite{walther2001carbonnanotubes}.
Such ordering allows closer packing of the water molecules leading
to a rise in density in the pore region.

\subsection{Voltage current relations}
\begin{figure*}
\includegraphics[scale=0.7]{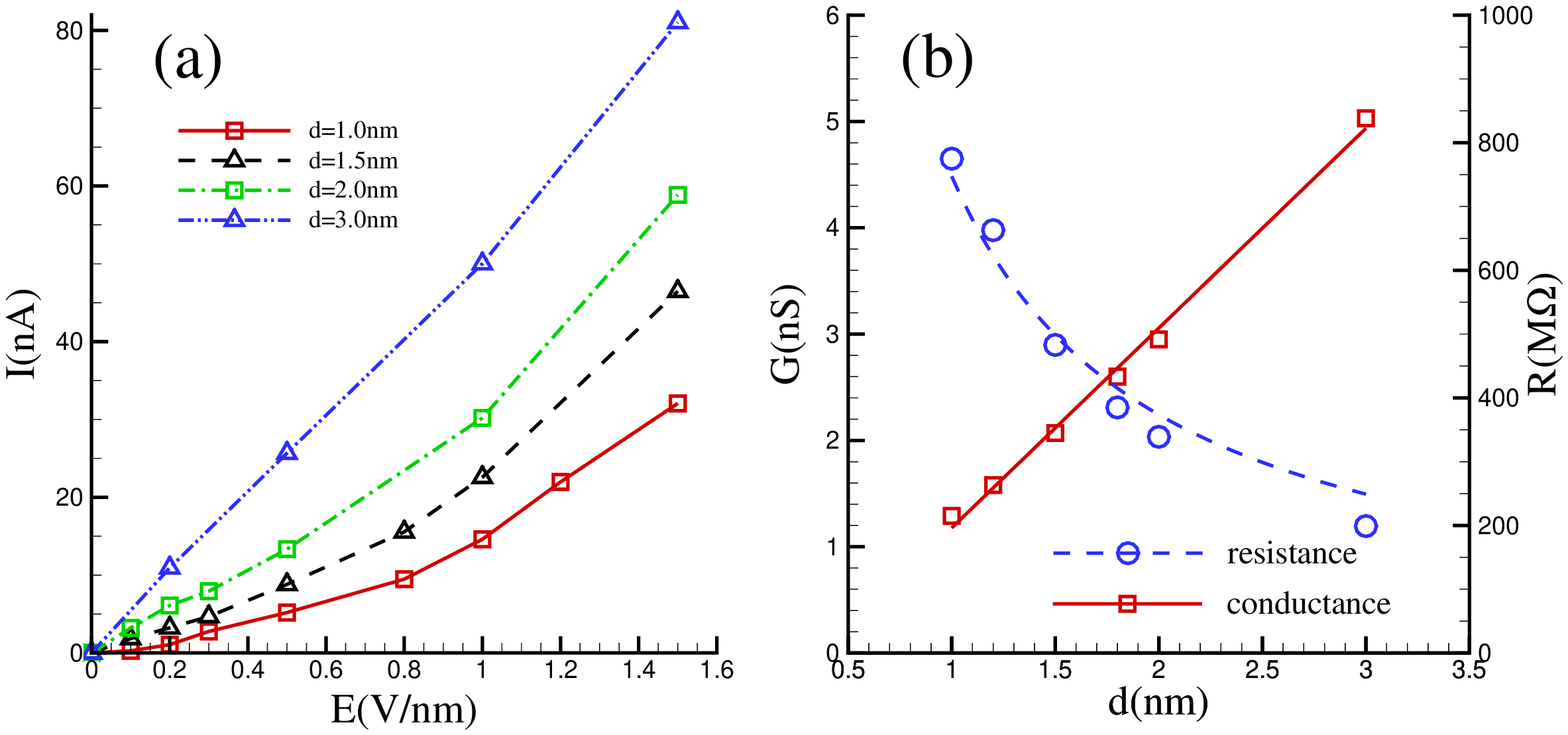} 
\caption{The current $vs$ applied field (left panel) for several different
pore diameters. The pore conductance determined from the slope of
the initial linear region of the current-voltage characteristic is
also shown as a function of the pore diameter (right panel). }
\label{current}
\end{figure*}
Recent experimental measurements have produced apparently conflicting
results on the dependence of pore conductance on pore radius. Garaj
et al. \cite{garaj2010graphene} reported that the pore conductance
is proportional to the pore diameter, whereas Schneider et al. \cite{schneider2010dnatranslocation}
reported it to be proportional to the pore area. The former result
would correspond to a hole in an infinitely thin membrane in a homogeneous
infinite conducting medium. The latter would correspond to a cylindrical
conduit of diameter much less than the length of the cylinder \cite{siwy2010nanopores}.
Since the former result is to be expected if the thickness of graphene
can be ignored and the electrolyte solution is assumed to obey Ohm's
law, some researchers \cite{venkatesan2011nanopore} have speculated
that the discrepancy between the experiment of Schneider et al. and
the theoretical result for a hole in a thin sheet might originate
from the fact that their graphene sheet was treated with a polymer
coating (to reduce non-specific surface adsorption of DNA). To clarify
this, Sathe et al. \cite{sathe2011computational} conducted molecular
dynamics simulation for a KCl solution with pore diameters in the
range of $2-7$ nm. They found the dependence of resistance $R$ on
pore diameter follows the relationship $R\sim1/d^{2}$, which is qualitatively
in agreement with the experiment of Schneider et al. \cite{schneider2010dnatranslocation}.
However, quantitatively the resistances they obtained in their simulations
were three to four times smaller than the experimental measurements.
They attributed the discrepancy to a number of factors including inaccuracies
in the force field in their simulations, unknown charge distribution
and uncertainties about the exact shape of graphene pores.

The variation of the current with the applied electric field was extracted
from our simulations and is shown in Fig. \ref{current} for a number of different
pore diameters, $d$. A linear regime is observed at weak fields ($E$).
For higher applied voltages, it is found  that the current shows super-linear
growth and the nonlinearity is stronger for pores of smaller diameter.
Hydrodynamic transport due to the nanoscale vortices could be responsible
for this faster than linear increase of the current. A similar phenomenon
has also been observed when an external electric field is applied
normal to the surface of an ion-selective nanoporous membrane immersed
in an electrolyte solution. There, at a critical value of the applied
field, the quiescent state undergoes a hydrodynamic instability resulting
in the appearance of micro vortices that are responsible for an ``overlimiting-current''
in the current-voltage relationship \cite{chang2012nanoscale}. In
the nanopore problem however, there is no instability. The hydrodynamic
flow is always present, its strength simply increases with the electric
field.

Fig. \ref{current}(b) shows the electric conductance ($G$) at low fields, which
is calculated from the slope of the initial linear segment of the
$I-V$ curves. It is seen to be proportional to the nanopore diameter
$d$, which is qualitatively in agreement with the experiments by
Garaj et al. \cite{garaj2010graphene} and is as expected from the
theoretical model where the pore is regarded as a hole in an infinitely
thin insulating sheet in a uniform conductor. The resistance obtained
for $d=3$ nm is $R=199$ M$\Omega$, which is comparable to the values
reported by Schneider et al \cite{schneider2010dnatranslocation},
though we do not observe their $G\propto d^{2}$ dependence on pore
diameter. The theoretical result $G=\sigma d$ for a hole of diameter
`$d$' in an insulating sheet in a conducting medium of conductivity
`$\sigma$' may be used to extract an ``effective'' conductivity
from the data shown in Fig. \ref{current}b. The conductivity obtained
in this way is $\sigma=1.879$ Sm$^{-1}$, which is about a factor
of $5$ smaller than the bulk conductivity, $\sigma=9.3$ Sm$^{-1}$,
for a 1 M NaCl solution. The discrepancy may arise from the fact that
there are significant differences between a graphene nanopore immersed
in an electrolyte and a hole in an insulating sheet in an Ohmic medium.
As we have seen, one important difference lies in the existence of
the concentration polarization layer which modifies the electric field
distribution near the pore as well as drives a hydrodynamic flow that
affects transport properties. The influence of these phenomena on
pore conductance is still an open problem. A second source of discrepancy
may be due to a limitation of our computation. In a physical experiment,
the nanopore radius is vanishingly small in comparison to the reservoir
dimensions, so that essentially all of the electrical resistance arises
from the pore. This assumption might not be satisfied as well in our
simulated model, since, due to limitations of computational capabilities,
the area of the graphene sheet that we constructed, $5nm\times5nm$,
is quite comparable to the pore diameter of $3$ nm. These issues
will be investigated further in future studies.

\section{Conclusion}

Molecular dynamics simulations were performed to study the transport
properties of an electrolyte through a nanometer sized pore. Due to
the partial blockage of the ionic current by the graphene membrane,
a concentration polarization layer (CPL) develops next to the membrane
when an external voltage is applied. The CPL is able to induce electric
pressure in the fluid adjacent to the pore. If the applied voltage
is large enough to overcome the effects of viscosity and Brownian
fluctuations, vortices are generated in the fluid near the pore. These
nanoscale vortices enhance the ionic current through a mechanism similar
to an effect responsible for the phenomenon of  ``overlimiting current'' for perm selective
membranes. Owing to the polarization of water molecules, the direction
of the electric field might have significant influence on water structure
near the graphene surface. This effect, together with the differing
mobility and van der Waals attraction to C exhibited by the Cl and
Na ions, brings about an asymmetric distribution of ions, and a net
flux of water in the direction of the electric field through the pore.
The average velocity of this hydrodynamic flow is found to increase
approximately linearly with the electric field and appears to be independent
of the pore diameter. Orientational order created in the water dipoles
by the electric field enhances closer packing of the water dipoles.
This is manifested in the appearance of regions of high water density
within the pore. The electrical conductance of the system is found
to vary linearly with the nanopore diameter, as one might expect from
the classical theoretical relation $G=\sigma d$ for a hole in an
insulating membrane within a conductor. However, the effective conductivity,
$G/d$ is found to be nearly 5 times the bulk conductivity of the
electrolyte. The linear dependence of $G$ on $d$ is in accord with
the experiments of Garaj et al. but we do not observe the $G\propto d^{2}$
dependence reported by Schneider et al. This could be due to differences
in experimental techniques as a result of which experimental conditions in the latter experiment do
not correspond to the model studied in this paper.

\begin{acknowledgement}

This work was supported by the National Science Foundation of China
(Grant No. 10872122), Science and Technology Commission of Shanghai
Municipality (Grant No. 10dz2212600), Research Fund for the Doctoral
Program of Higher Education of China (Grant No. 20103108110004), and
Shanghai Program for Innovative Research Team in Universities. This
project was supported in part by the American Recovery and Reinvestment
Act (ARRA) funds through grant number R01 HG001234 to Northwestern
University (USA) from the National Human Genome Research Institute,
National Institutes of Health.
\end{acknowledgement}
\bibliographystyle{achemso.bst}
%\bibliography{nanopore}

\begin{mcitethebibliography}{37}
\providecommand*\natexlab[1]{#1}
\providecommand*\mciteSetBstSublistMode[1]{}
\providecommand*\mciteSetBstMaxWidthForm[2]{}
\providecommand*\mciteBstWouldAddEndPuncttrue
  {\def\EndOfBibitem{\unskip.}}
\providecommand*\mciteBstWouldAddEndPunctfalse
  {\let\EndOfBibitem\relax}
\providecommand*\mciteSetBstMidEndSepPunct[3]{}
\providecommand*\mciteSetBstSublistLabelBeginEnd[3]{}
\providecommand*\EndOfBibitem{}
\mciteSetBstSublistMode{f}
\mciteSetBstMaxWidthForm{subitem}{(\alph{mcitesubitemcount})}
\mciteSetBstSublistLabelBeginEnd
  {\mcitemaxwidthsubitemform\space}
  {\relax}
  {\relax}

\bibitem[Doyle et~al.(1998)Doyle, Cabral, Pfuetzner, Kuo, Gulbis, Cohen, Chait,
  and {MacKinnon}]{doyle1998thestructure}
Doyle,~D.~A.; Cabral,~J.~a.~M.; Pfuetzner,~R.~A.; Kuo,~A.; Gulbis,~J.~M.;
  Cohen,~S.~L.; Chait,~B.~T.; {MacKinnon},~R. \emph{Science} \textbf{1998},
  \emph{280}, 69 --77\relax
\mciteBstWouldAddEndPuncttrue
\mciteSetBstMidEndSepPunct{\mcitedefaultmidpunct}
{\mcitedefaultendpunct}{\mcitedefaultseppunct}\relax
\EndOfBibitem
\bibitem[Rhee and Burns(2006)Rhee, and Burns]{rhee2006nanopore}
Rhee,~M.; Burns,~M.~A. \emph{Trends in Biotechnology} \textbf{2006}, \emph{24},
  580--586\relax
\mciteBstWouldAddEndPuncttrue
\mciteSetBstMidEndSepPunct{\mcitedefaultmidpunct}
{\mcitedefaultendpunct}{\mcitedefaultseppunct}\relax
\EndOfBibitem
\bibitem[Venkatesan and Bashir(2011)Venkatesan, and
  Bashir]{venkatesan2011nanopore}
Venkatesan,~B.~M.; Bashir,~R. \emph{Nat Nano} \textbf{2011}, \emph{6},
  615--624\relax
\mciteBstWouldAddEndPuncttrue
\mciteSetBstMidEndSepPunct{\mcitedefaultmidpunct}
{\mcitedefaultendpunct}{\mcitedefaultseppunct}\relax
\EndOfBibitem
\bibitem[Branton et~al.(2008)Branton, Deamer, Marziali, Bayley, Benner, Butler,
  Di~Ventra, Garaj, Hibbs, Huang, Jovanovich, Krstic, Lindsay, Ling,
  Mastrangelo, Meller, Oliver, Pershin, Ramsey, Riehn, Soni, {Tabard-Cossa},
  Wanunu, Wiggin, and Schloss]{branton2008thepotential}
Branton,~D. et~al.  \emph{Nat Biotech} \textbf{2008}, \emph{26},
  1146--1153\relax
\mciteBstWouldAddEndPuncttrue
\mciteSetBstMidEndSepPunct{\mcitedefaultmidpunct}
{\mcitedefaultendpunct}{\mcitedefaultseppunct}\relax
\EndOfBibitem
\bibitem[Roux et~al.(2004)Roux, Allen, Bern\`{e}che, and
  Im]{roux2004theoretical}
Roux,~B.; Allen,~T.; Bern\`{e}che,~S.; Im,~W. \emph{Quarterly Reviews of
  Biophysics} \textbf{2004}, \emph{37}, 15--103\relax
\mciteBstWouldAddEndPuncttrue
\mciteSetBstMidEndSepPunct{\mcitedefaultmidpunct}
{\mcitedefaultendpunct}{\mcitedefaultseppunct}\relax
\EndOfBibitem
\bibitem[Aksimentiev et~al.(2009)Aksimentiev, Brunner, {Cruz-Chu}, Comer, and
  Schulten]{aksimentiev2009modeling}
Aksimentiev,~A.; Brunner,~R.; {Cruz-Chu},~E.; Comer,~J.; Schulten,~K.
  \emph{{IEEE} Nanotechnology Magazine} \textbf{2009}, \emph{3}, 20--28\relax
\mciteBstWouldAddEndPuncttrue
\mciteSetBstMidEndSepPunct{\mcitedefaultmidpunct}
{\mcitedefaultendpunct}{\mcitedefaultseppunct}\relax
\EndOfBibitem
\bibitem[Aksimentiev and Schulten(2005)Aksimentiev, and
  Schulten]{aksimentiev2005imaging}
Aksimentiev,~A.; Schulten,~K. \emph{Biophysical Journal} \textbf{2005},
  \emph{88}, 3745--3761\relax
\mciteBstWouldAddEndPuncttrue
\mciteSetBstMidEndSepPunct{\mcitedefaultmidpunct}
{\mcitedefaultendpunct}{\mcitedefaultseppunct}\relax
\EndOfBibitem
\bibitem[Comer et~al.(2009)Comer, Dimitrov, Zhao, Timp, and
  Aksimentiev]{comer2009microscopic}
Comer,~J.; Dimitrov,~V.; Zhao,~Q.; Timp,~G.; Aksimentiev,~A. \emph{Biophysical
  Journal} \textbf{2009}, \emph{96}, 593--608\relax
\mciteBstWouldAddEndPuncttrue
\mciteSetBstMidEndSepPunct{\mcitedefaultmidpunct}
{\mcitedefaultendpunct}{\mcitedefaultseppunct}\relax
\EndOfBibitem
\bibitem[{Cruz-Chu} et~al.(2009){Cruz-Chu}, Aksimentiev, and
  Schulten]{cruz-chu2009ioniccurrent}
{Cruz-Chu},~E.~R.; Aksimentiev,~A.; Schulten,~K. \emph{The Journal of Physical
  Chemistry. C, Nanomaterials and Interfaces} \textbf{2009}, \emph{113}, 1850,
  {PMID:} 20126282\relax
\mciteBstWouldAddEndPuncttrue
\mciteSetBstMidEndSepPunct{\mcitedefaultmidpunct}
{\mcitedefaultendpunct}{\mcitedefaultseppunct}\relax
\EndOfBibitem
\bibitem[Sathe et~al.(2011)Sathe, Zou, Leburton, and
  Schulten]{sathe2011computational}
Sathe,~C.; Zou,~X.; Leburton,~J.; Schulten,~K. \emph{{ACS} Nano} \textbf{2011},
  \emph{5}, 8842--8851, {PMID:} 21981556\relax
\mciteBstWouldAddEndPuncttrue
\mciteSetBstMidEndSepPunct{\mcitedefaultmidpunct}
{\mcitedefaultendpunct}{\mcitedefaultseppunct}\relax
\EndOfBibitem
\bibitem[Suk and Aluru(2010)Suk, and Aluru]{suk2010watertransport}
Suk,~M.~E.; Aluru,~N.~R. \emph{J. Phys. Chem. Lett.} \textbf{2010}, \emph{1},
  1590--1594\relax
\mciteBstWouldAddEndPuncttrue
\mciteSetBstMidEndSepPunct{\mcitedefaultmidpunct}
{\mcitedefaultendpunct}{\mcitedefaultseppunct}\relax
\EndOfBibitem
\bibitem[Aksimentiev et~al.(2004)Aksimentiev, Heng, Timp, and
  Schulten]{aksimentiev2004microscopic}
Aksimentiev,~A.; Heng,~J.~B.; Timp,~G.; Schulten,~K. \emph{Biophysical Journal}
  \textbf{2004}, \emph{87}, 2086--2097\relax
\mciteBstWouldAddEndPuncttrue
\mciteSetBstMidEndSepPunct{\mcitedefaultmidpunct}
{\mcitedefaultendpunct}{\mcitedefaultseppunct}\relax
\EndOfBibitem
\bibitem[Luan and Aksimentiev(2010)Luan, and Aksimentiev]{luan2010control}
Luan,~B.; Aksimentiev,~A. \emph{Journal of Physics: Condensed Matter}
  \textbf{2010}, \emph{22}, 454123\relax
\mciteBstWouldAddEndPuncttrue
\mciteSetBstMidEndSepPunct{\mcitedefaultmidpunct}
{\mcitedefaultendpunct}{\mcitedefaultseppunct}\relax
\EndOfBibitem
\bibitem[Mirsaidov et~al.(2010)Mirsaidov, Comer, Dimitrov, Aksimentiev, and
  Timp]{mirsaidov2010slowing}
Mirsaidov,~U.; Comer,~J.; Dimitrov,~V.; Aksimentiev,~A.; Timp,~G.
  \emph{Nanotechnology} \textbf{2010}, \emph{21}, 395501\relax
\mciteBstWouldAddEndPuncttrue
\mciteSetBstMidEndSepPunct{\mcitedefaultmidpunct}
{\mcitedefaultendpunct}{\mcitedefaultseppunct}\relax
\EndOfBibitem
\bibitem[Ghosal(2006)]{ghosal_PRE06}
Ghosal,~S. \emph{Phys. Rev. E} \textbf{2006}, \emph{74},
  041901--1--041901--5\relax
\mciteBstWouldAddEndPuncttrue
\mciteSetBstMidEndSepPunct{\mcitedefaultmidpunct}
{\mcitedefaultendpunct}{\mcitedefaultseppunct}\relax
\EndOfBibitem
\bibitem[Ghosal(2007)]{ghosal_PRL07}
Ghosal,~S. \emph{Phys. Rev. Lett.} \textbf{2007}, \emph{98}, 238104\relax
\mciteBstWouldAddEndPuncttrue
\mciteSetBstMidEndSepPunct{\mcitedefaultmidpunct}
{\mcitedefaultendpunct}{\mcitedefaultseppunct}\relax
\EndOfBibitem
\bibitem[Storm et~al.(2005)Storm, Chen, Zandbergen, and
  Dekker]{storm_physRevE05}
Storm,~A.; Chen,~J.; Zandbergen,~H.; Dekker,~C. \emph{Phys. Rev. E}
  \textbf{2005}, \emph{71}, 051903--1--051903--10\relax
\mciteBstWouldAddEndPuncttrue
\mciteSetBstMidEndSepPunct{\mcitedefaultmidpunct}
{\mcitedefaultendpunct}{\mcitedefaultseppunct}\relax
\EndOfBibitem
\bibitem[Smeets et~al.(2006)Smeets, Keyser, Krapf, Wu, N.H., and
  C.]{dekker_nano_lett06}
Smeets,~M.; Keyser,~U.; Krapf,~D.; Wu,~M.; N.H.,~D.; C.,~D. \emph{Nano Letters}
  \textbf{2006}, \emph{6}, 89--95\relax
\mciteBstWouldAddEndPuncttrue
\mciteSetBstMidEndSepPunct{\mcitedefaultmidpunct}
{\mcitedefaultendpunct}{\mcitedefaultseppunct}\relax
\EndOfBibitem
\bibitem[van Dorp et~al.(2009)van Dorp, Keyser, Dekker, Dekker, and
  Lemay]{van_dorp_origin_2009}
van Dorp,~S.; Keyser,~U.~F.; Dekker,~N.~H.; Dekker,~C.; Lemay,~S.~G. \emph{Nat
  Phys} \textbf{2009}, \emph{5}, 347--351\relax
\mciteBstWouldAddEndPuncttrue
\mciteSetBstMidEndSepPunct{\mcitedefaultmidpunct}
{\mcitedefaultendpunct}{\mcitedefaultseppunct}\relax
\EndOfBibitem
\bibitem[Ghosal(2007)]{ghosal_PRE07}
Ghosal,~S. \emph{Phys. Rev. E} \textbf{2007}, \emph{76}, 061916\relax
\mciteBstWouldAddEndPuncttrue
\mciteSetBstMidEndSepPunct{\mcitedefaultmidpunct}
{\mcitedefaultendpunct}{\mcitedefaultseppunct}\relax
\EndOfBibitem
\bibitem[Chang et~al.(2012)Chang, Yossifon, and Demekhin]{chang2012nanoscale}
Chang,~H.; Yossifon,~G.; Demekhin,~E.~A. \emph{Annual Review of Fluid
  Mechanics} \textbf{2012}, \emph{44}, 401--426\relax
\mciteBstWouldAddEndPuncttrue
\mciteSetBstMidEndSepPunct{\mcitedefaultmidpunct}
{\mcitedefaultendpunct}{\mcitedefaultseppunct}\relax
\EndOfBibitem
\bibitem[Berendsen et~al.(1987)Berendsen, Grigera, and
  Straatsma]{berendsen1987themissing}
Berendsen,~H. J.~C.; Grigera,~J.~R.; Straatsma,~T.~P. \emph{J. Phys. Chem.}
  \textbf{1987}, \emph{91}, 6269--6271\relax
\mciteBstWouldAddEndPuncttrue
\mciteSetBstMidEndSepPunct{\mcitedefaultmidpunct}
{\mcitedefaultendpunct}{\mcitedefaultseppunct}\relax
\EndOfBibitem
\bibitem[Yang et~al.(2002)Yang, Yiacoumi, and Tsouris]{yang2002canonical}
Yang,~K.; Yiacoumi,~S.; Tsouris,~C. \emph{The Journal of Chemical Physics}
  \textbf{2002}, \emph{117}, 337--345\relax
\mciteBstWouldAddEndPuncttrue
\mciteSetBstMidEndSepPunct{\mcitedefaultmidpunct}
{\mcitedefaultendpunct}{\mcitedefaultseppunct}\relax
\EndOfBibitem
\bibitem[Hess et~al.(2008)Hess, Kutzner, van~der Spoel, and
  Lindahl]{hess2008gromacs}
Hess,~B.; Kutzner,~C.; van~der Spoel,~D.; Lindahl,~E. \emph{J. Chem. Theory
  Comput.} \textbf{2008}, \emph{4}, 435--447\relax
\mciteBstWouldAddEndPuncttrue
\mciteSetBstMidEndSepPunct{\mcitedefaultmidpunct}
{\mcitedefaultendpunct}{\mcitedefaultseppunct}\relax
\EndOfBibitem
\bibitem[Gong et~al.(2007)Gong, Li, Lu, Wan, Li, Hu, and
  Fang]{gong2007achargedriven}
Gong,~X.; Li,~J.; Lu,~H.; Wan,~R.; Li,~J.; Hu,~J.; Fang,~H. \emph{Nat Nano}
  \textbf{2007}, \emph{2}, 709--712\relax
\mciteBstWouldAddEndPuncttrue
\mciteSetBstMidEndSepPunct{\mcitedefaultmidpunct}
{\mcitedefaultendpunct}{\mcitedefaultseppunct}\relax
\EndOfBibitem
\bibitem[Hummer et~al.(2001)Hummer, Rasaiah, and
  Noworyta]{hummer2001waterconduction}
Hummer,~G.; Rasaiah,~J.~C.; Noworyta,~J.~P. \emph{Nature} \textbf{2001},
  \emph{414}, 188--190\relax
\mciteBstWouldAddEndPuncttrue
\mciteSetBstMidEndSepPunct{\mcitedefaultmidpunct}
{\mcitedefaultendpunct}{\mcitedefaultseppunct}\relax
\EndOfBibitem
\bibitem[{Xiao-Yan} and {Hang-Jun}(2007){Xiao-Yan}, and
  {Hang-Jun}]{xiao-yan2007thestructure}
{Xiao-Yan},~Z.; {Hang-Jun},~L. \emph{Chinese Physics} \textbf{2007}, \emph{16},
  335--339\relax
\mciteBstWouldAddEndPuncttrue
\mciteSetBstMidEndSepPunct{\mcitedefaultmidpunct}
{\mcitedefaultendpunct}{\mcitedefaultseppunct}\relax
\EndOfBibitem
\bibitem[Darden et~al.(1993)Darden, York, and Pedersen]{darden1993particle}
Darden,~T.; York,~D.; Pedersen,~L. \emph{The Journal of Chemical Physics}
  \textbf{1993}, \emph{98}, 10089\relax
\mciteBstWouldAddEndPuncttrue
\mciteSetBstMidEndSepPunct{\mcitedefaultmidpunct}
{\mcitedefaultendpunct}{\mcitedefaultseppunct}\relax
\EndOfBibitem
\bibitem[Bratko et~al.(2007)Bratko, Daub, Leung, and Luzar]{bratko2007effectof}
Bratko,~D.; Daub,~C.~D.; Leung,~K.; Luzar,~A. \emph{J. Am. Chem. Soc.}
  \textbf{2007}, \emph{129}, 2504--2510\relax
\mciteBstWouldAddEndPuncttrue
\mciteSetBstMidEndSepPunct{\mcitedefaultmidpunct}
{\mcitedefaultendpunct}{\mcitedefaultseppunct}\relax
\EndOfBibitem
\bibitem[Daub et~al.(2006)Daub, Bratko, Leung, and
  Luzar]{daub2006electrowetting}
Daub,~C.~D.; Bratko,~D.; Leung,~K.; Luzar,~A. \emph{J. Phys. Chem. C}
  \textbf{2006}, \emph{111}, 505--509\relax
\mciteBstWouldAddEndPuncttrue
\mciteSetBstMidEndSepPunct{\mcitedefaultmidpunct}
{\mcitedefaultendpunct}{\mcitedefaultseppunct}\relax
\EndOfBibitem
\bibitem[Daub et~al.(2012)Daub, Bratko, and Luzar]{daub2012nanoscale}
Daub,~C.~D.; Bratko,~D.; Luzar,~A. \emph{Topics in Current Chemistry}
  \textbf{2012}, \emph{307}, 155--179\relax
\mciteBstWouldAddEndPuncttrue
\mciteSetBstMidEndSepPunct{\mcitedefaultmidpunct}
{\mcitedefaultendpunct}{\mcitedefaultseppunct}\relax
\EndOfBibitem
\bibitem[Hu et~al.(2008)Hu, Xu, Xu, and Zhou]{hu2008dewetting}
Hu,~G.; Xu,~A.; Xu,~Z.; Zhou,~Z. \emph{Physics of Fluids} \textbf{2008},
  \emph{20}, 102101\relax
\mciteBstWouldAddEndPuncttrue
\mciteSetBstMidEndSepPunct{\mcitedefaultmidpunct}
{\mcitedefaultendpunct}{\mcitedefaultseppunct}\relax
\EndOfBibitem
\bibitem[Walther et~al.(2001)Walther, Jaffe, Halicioglu, and
  Koumoutsakos]{walther2001carbonnanotubes}
Walther,~J.~H.; Jaffe,~R.; Halicioglu,~T.; Koumoutsakos,~P. \emph{J. Phys.
  Chem. B} \textbf{2001}, \emph{105}, 9980--9987\relax
\mciteBstWouldAddEndPuncttrue
\mciteSetBstMidEndSepPunct{\mcitedefaultmidpunct}
{\mcitedefaultendpunct}{\mcitedefaultseppunct}\relax
\EndOfBibitem
\bibitem[Garaj et~al.(2010)Garaj, Hubbard, Reina, Kong, Branton, and
  Golovchenko]{garaj2010graphene}
Garaj,~S.; Hubbard,~W.; Reina,~A.; Kong,~J.; Branton,~D.; Golovchenko,~J.~A.
  \emph{Nature} \textbf{2010}, \emph{467}, 190--193\relax
\mciteBstWouldAddEndPuncttrue
\mciteSetBstMidEndSepPunct{\mcitedefaultmidpunct}
{\mcitedefaultendpunct}{\mcitedefaultseppunct}\relax
\EndOfBibitem
\bibitem[Schneider et~al.(2010)Schneider, Kowalczyk, Calado, Pandraud,
  Zandbergen, Vandersypen, and Dekker]{schneider2010dnatranslocation}
Schneider,~G.~F.; Kowalczyk,~S.~W.; Calado,~V.~E.; Pandraud,~G.;
  Zandbergen,~H.~W.; Vandersypen,~L. M.~K.; Dekker,~C. \emph{Nano Lett.}
  \textbf{2010}, \emph{10}, 3163--3167\relax
\mciteBstWouldAddEndPuncttrue
\mciteSetBstMidEndSepPunct{\mcitedefaultmidpunct}
{\mcitedefaultendpunct}{\mcitedefaultseppunct}\relax
\EndOfBibitem
\bibitem[Siwy and Davenport(2010)Siwy, and Davenport]{siwy2010nanopores}
Siwy,~Z.~S.; Davenport,~M. \emph{Nat Nano} \textbf{2010}, \emph{5},
  697--698\relax
\mciteBstWouldAddEndPuncttrue
\mciteSetBstMidEndSepPunct{\mcitedefaultmidpunct}
{\mcitedefaultendpunct}{\mcitedefaultseppunct}\relax
\EndOfBibitem
\end{mcitethebibliography}
\providecommand*\mcitethebibliography{\thebibliography}
\csname @ifundefined\endcsname{endmcitethebibliography}
  {\let\endmcitethebibliography\endthebibliography}{}

\end{document}